\documentstyle[12pt,amstex,amsfonts,amssymb,rotating,epsf]{article}

\begin{document}
 
\title{{\bf Characterization of the long-time and short-time
  predictability of low-order models of the atmosphere}}
  \author{ R. Benzi$^{1}$, M. Marrocu$^{2}$, A. Mazzino$^{3}$
  and E. Trovatore$^{4}$ }
\date{}
  \maketitle
  \centerline{$^{1}$  Autorit\`a per l'Informatica nella Pubblica 
  Amministrazione,}
  \centerline{Via Solferino 15, I-00100 Roma, Italy.}
  \centerline{$^{2}$  CRS4, C.P. 94, I-09010 Uta (Cagliari), Italy.}
  \centerline{$^{3}$  INFM--Dipartimento di Fisica, Universit\`{a}
di Genova,}
  \centerline{Via Dodecaneso 33, I-16146, Genova, Italy.}
  \centerline{$^{4}$ Centro Meteo--Idrologico della Regione Liguria, }
  \centerline{Via Dodecaneso 33, I-16146, Genova, Italy.}
\vspace{0.2cm}


\vspace{0.2cm}

 \medskip

\begin{abstract}
Methods to quantify predictability properties
of atmospheric flows are proposed.  
The ``Extended Self Similarity'' (ESS) technique,
recently employed  in turbulence data analysis,
is used
to characterize predictability properties at short
and long times.  
We apply our methods to the 
low-order  
atmospheric model
of Lorenz 
(1980).
We also investigate how 
initialization procedures 
that eliminate gravity waves from the model dynamics
influence predictability properties.
\end{abstract}


\section{Introduction}
\label{sec:intro}
The behavior of a system such as the atmosphere
cannot be predicted with prescribed  spatial
detail beyond a certain characteristic time 
because initially close trajectories diverge with time.
Since the first systematic studies of error growth,
low--order models of the atmospheric circulation, such as the
Lorenz (1963,1980) models, have been used to study the predictability
properties of chaotic systems in general and of the atmosphere in
particular.
In spite of the their lack of realism,
low order models can be still considered useful tools to study some of
the mechanisms
that limit predictability.
Several different methods have been adopted in studies on the
subject.
In this paper we will follow the guidelines of 
methods employed in the theory of dynamical
systems (Benzi {\em et al.} 1985; Benzi and Carnevale 1989) 
in order to study the statistics of divergence  of initially
close forecasts.  

As is well-known (Oseledec 1968), 
for chaotic systems
the asymptotic divergence of initially close trajectories behave
exponentially
with a rate
given by the maximum Lyapunov exponent $\lambda_1$ 
(Benettin {\em et al.} 1980). 
Lyapunov exponents are defined as global averages on the attractor of
a chaotic system;
however, predictability is a local property in the phase space. 
Indeed, operational forecasting experience indicates that the skill
associated
with individual forecasts may vary greatly.
Thus, predictability properties of atmospheric models exhibit large
variability in both time and phase space (Palmer et al. 1990);
that is, predictability is characterized by the presence of
fluctuations.

Another important point is the existence of a 
period of transient growth that can exceed the Lyapunov exponential
growth
rate (Mukougawa {\em et al.} 1991; 
Nicolis 1992; Trevisan 1993; Trevisan and Legnani 1995). 
Error amplification on short time scales is controlled by rapidly
growing perturbations that are not of normal-mode form. If the normal
modes of instability of a stationary solution are not orthogonal,
perturbations may for some time grow faster than the most unstable
normal mode and the global average growth rate of errors, initially in a random
direction, attains the asymptotic value given by the first Lyapunov
exponent after a finite period of time. The concept of transient
enhanced exponential growth was introduced by Farrell (1985) and
Lacarra and Talagrand (1988). 
A common application  of this concept used in numerical weather
prediction consists in finding the initial perturbation that grows
fastest in a given period of time (ECMWF 1992; Mureau {\it et al.}
1993;
Molteni {\it et al.} 1996). 

In this paper we want to address the following
questions:
\begin{itemize}
\item how can the fluctuations in the predictability properties
of atmospheric models be described?
\item how can the predictability properties in the transient
region of super-exponential growth be described?
\item what is the influence of initialization procedures on the
predictability fluctuations? 
\end{itemize}

Our major goal  will be to give new insights on
these three (related) problems. Specifically, we will use, for the
first time in
atmospheric science, mathematical 
tools which are well-known in the theory of dynamical
systems and in other contexts (as in turbulence theory) 
and which  allow the quantitative characterization of predictability
fluctuations at long and short times.

The outline of the paper is as follows:
in  sec.~2, we discuss the mathematical background needed to
describe the statistics of error growth.
A quantitative definition of its statistical
fluctuations (a characteristic also known as {\it intermittency})
is given in
terms of an infinite set of characteristic time scales, related to 
the so-called generalized Lyapunov exponents $L(q)$.
In the absence of fluctuations, the generalized
Lyapunov exponents are linearly dependent on 
$\lambda_1$, thus making the maximum Lyapunov exponent the  
only relevant parameter defining the predictability.
We also discuss how to characterize the predictability fluctuations at
short times by means of new techniques recently applied in turbulence
theory.
In sec.~3, we apply these ideas to 
the
low-order model introduced and studied by Lorenz (1980).
In sec.~4 we investigate how predictability properties are
modified  by the superbalance equations introduced by Lorenz (1980), 
which do not not support gravity wave oscillations. This allows us
to make an
assessment of the role played by gravity wave oscillations in
enhancing predictability fluctuations at short times.
Conclusions follow  in sec.~\ref{sec:conclu}.

\section{Predictability fluctuations: the general theory}
\label{sec:theory}

In this section we shall review some basic mathematical definitions 
and concepts related to sensitive dependence of the trajectory of a
generic
dynamical system on initial conditions. 

Let us consider an N-dimensional dynamical system given by the
equations:

\begin{equation}
{\dot{\bf{x}}}={\bf{f}}({\bf{x}}),
\label{eq:dyn}
\end{equation}
where ${\bf{x}}\in R^N$. The time evolution of an infinitesimal
disturbance $\delta{\bf{x}}(t)$ is given by the linearized equations:

\begin{equation}
\delta{\dot{\bf{x}}}(t)={\bf{Df}} \,\, \delta{\bf{x}}(t),
\label{eq:dyn_error}
\end{equation}
where $\delta{\bf{x}}(0) \neq 0$ and ${D f}_{i j}=
\left.\displaystyle{\frac{\partial f_i}{\partial x_j}} \right| 
_{{\bf{x}}={\bf{x}}(t)}$.

We can define the response function $R(t,0)$ as

\begin{equation}
R(t,0)=\frac{|\delta{\bf{x}}(t)|} { |\delta{\bf{x}}(0)|}.
\label{eq:g0}
\end{equation}

The Oseledec theorem (Oseledec 1968)
tells us that, for $t \rightarrow \infty$
and for almost all (in the sense of measure theory) initial conditions
$\delta{\bf{x}}(0)$, we have:

\begin{equation}
R(t,0) \rightarrow e^{\lambda_1 t},
\label{eq:g1}
\end{equation}
where $\lambda_1$ is called the {\it maximum Lyapunov exponent}
(Benettin 
{\em et al.} 1980). 
One may venture to guess that this exponential growth is valid even
at finite times and that $\lambda_1^{-1}$ is the only characteristic
time scale of the error growth;
however as is well-known, this is not the case (see, for example,
Trevisan and
Legnani 1995).
Indeed, even 
if $t$ is large enough to allow the
observation of
an exponential growth rate of $R(t,0)$,
 $\lambda_1^{-1}$ is not the only relevant time scale:
depending on the initial conditions, 
dynamical systems can show
states or configurations 
that can be predicted for times longer or shorter than
$\lambda_1^{-1}$.

The question we want to address is related to
 the existence of well-defined  mathematical quantities able to
characterize
these  predictability fluctuations in 
chaotic systems (Eckmann and Procaccia 1986;
Paladin {\em et al.} 1986) and in particular in atmospheric models
 (Benzi and Carnevale 1989; Trevisan and Legnani 1995). 
This aspect will be investigated in sec.~\ref{sec:large_t}.

Another 
issue to address is that
for times not long enough, the dynamical behavior of $R(t,0)$
is not necessarily characterized by an exponential growth rate. This
question
has been investigated 
by Mukougawa {\em et al.} (1991), 
Nicolis (1992), and Trevisan (1993), 
Trevisan and Legnani (1995), amongst others,
and turns out to be of great importance in
realistic applications of predictability theory.
We shall come to this question in sec.~\ref{sec:short_t}.

\subsection{Characterization of predictability fluctuations at long times}
\label{sec:large_t}

In this section we recall the concept of  generalized Lyapunov
exponents,
already introduced in atmospheric applications by Benzi and Carnevale
(1989).
We point out that $\lambda_1$ is an average quantity 
and we define the probability $P_t(\gamma)$ of having a local exponent
$\gamma$
different from $\lambda_1$. Then we derive the
relationships between all these quantities and we give two examples of
simple probability density functions (p.d.f.'s).
In particular, we show that the log-Poisson p.d.f, proposed here for
the
first time in a meteorological context, has some
interesting features regarding the description of strong fluctuations.
  
\subsubsection{Basic concepts}

In this section we review some general statistical properties of the
response function $R(t,0)$ (see also Benzi and Carnevale 1989). 
Let us consider times long enough  to have an exponential
growth rate for the response function $R(t,0)$. 
Let us introduce the quantities:

\begin{equation}
 \langle R(t,0)^q \rangle,
\end{equation}
where $\langle \cdot \rangle$ is the average over different initial
conditions.

If $\lambda_1$ were the only relevant time scale characterizing 
the error growth,
then we should expect:

\begin{equation}
 \langle R(t,0)^q \rangle \sim e^{\lambda_1 q t}.
\label{eq:no_int}
\end{equation}

On the other hand, if many times scales  characterize the error
growth, we
should have the more general behavior:

\begin{equation}
 \langle R(t,0)^q \rangle \sim e^{L(q) t},
\label{eq:int}
\end{equation}
where the $L(q)$'s are
the so-called {\it generalized Lyapunov exponents}
(Fujisaka 1983; Benzi {\em et al.} 1985). 
In general,
$L(q)$ is not a  linear function of $q$.

Before explaining in detail the physical meaning of
eq.~(\ref{eq:int}),
let us give simple examples of  dynamical systems that
satisfy either eq.~(\ref{eq:no_int}) or eq.~(\ref{eq:int}).

Let us consider the one dimensional ``tent'' map (see Fig.~1):

\begin{equation}
x_{n+1} = \left\{
\begin{array}{ll}
\frac{x_n}{c} \,\,\,\,\,\,\,\,\,\,\, &(0 \le x \le c) \\
\frac{1-x_n}{1-c}\,\,\,\,\,\,\,\,\,\,\,\,\,\,\, &(c < x \le 1) .\\
\end{array}
\right .
\label{eq:map}
\end{equation}
In this case, $t=n$ is a discrete time and the corresponding response 
function $R(n,0)$ is given by:

\begin{equation}
 R(n,0)= \prod_{i=1}^n D_i,
\label{eq:Rmap}
\end{equation}
where $D_i$ is given either by $1/c$ or $1/(1-c)$, depending on
whether $x_i \in [0,c]$ or $x_i \in [c,1]$, respectively.

In order to compute $\langle R(n,0)^q \rangle$, 
we need to know the p.d.f. of $x$.
For the tent map the p.d.f. is that of the uniform distribution on the
interval $[0,1]$ (Frisch 1995).
It then follows from (\ref{eq:Rmap}) that:

\begin{equation}\begin{array}{ll}
\langle R(n,0)^q \rangle &=  \{ \int_0^c [\frac{1}{c}]^q dx +
                \int_c^1 [\frac{1}{1-c}]^q dx \}^n \\
          &=  \{ c^{1-q}+(1-c)^{1-q} \}^n = e^{L(q) n},
\end{array}
\label{eq:mom_map}
\end{equation}
where 

\begin{equation}
 L(q)=\ln [c^{1-q}+(1-c)^{1-q}].
\label{eq:Lmap}
\end{equation}

First of all, notice that for $c=1/2$, all points of the 
dynamical system are characterized by the same slope $D_n = 2$.
In this case we have:

\begin{equation}
 L(q)=\ln \left[ \left( {\frac{1}{2}}\right )^{1-q}
+{\left( \frac{1}{2} \right ) }^{1-q}\right ]=q \, ln 2 \;\;\; .
\label{eq:Lmap2}
\end{equation}
Secondly, for $c>1/2$ (and in particular for $c$ very close to 1)
the system  is characterized by two different slopes, 
$1/c < 2$ and $1/(1-c) >2$;
this means that there are states of the system for which the error
growth
is `slow' ($x \in [0,c]$) and states for which the error growth is
`fast'
($x \in [c,1]$).
In this case, $L(q)$ is no longer characterized by a single time
scale.

Coming back to the interpretation of eq.~(\ref{eq:int}) and 
assuming that
$R(t,0)$ already shows an exponential behavior, we can write:

\begin{equation}
R(t,0)\sim e^{\gamma t},
\label{eq:gamma}
\end{equation}
where $\gamma$ is the {\it local} error growth exponent; the 
error growth exponent is local in the sense
that
it depends on 
the particular
initial condition under consideration.
We now briefly review how $\lambda_1$ can be seen as the average over
all possible local exponents of the system.
We can rewrite $R(t,0)$ as:

\begin{equation}
R(t,0)= \prod_{i=1}^M R(t_i,t_{i-1}),
\label{eq:Ri}
\end{equation}
where $0=t_0<t_1<...<t_i<t_{i+1}<...<t_M=t$, and we are regarding the
trajectory
as a sequence of $M$ trajectories. 
By introducing the notation:

\begin{equation}
R(t_i,t_{i-1})\sim e^{\gamma_i (t_i-t_{i-1})},
\label{eq:gammai}
\end{equation}
we have:

\begin{equation}
e^{\gamma t}= e^{\sum_{i=1}^M \gamma_i (t_i-t_{i-1})}.
\label{eq:rel}
\end{equation}
Therefore, assuming $t_i-t_{i-1}=\Delta t$ for any $i$, we have $t=M
\Delta t$
and

\begin{equation}
\gamma = \frac{1}{M} \sum_{i=1}^M \gamma_i.
\label{eq:sum}
\end{equation}
Equation (\ref{eq:sum}) tells us that $\gamma$ is just the average
value of 
$\gamma_i$'s along the trajectory.
By the Oseledec (1968) theorem we know that, for $t\rightarrow \infty$
(i.e., for $M\rightarrow \infty$), $\gamma \rightarrow \lambda_1 $.
Thus, $\lambda_1$ is the average over
all  possible local exponents $\gamma_i$ of the system.    
We can now ask what is the probability of having
$\gamma \ne \lambda_1$ for finite times.
In order to answer this question, we introduce  the
probability $P_t(\gamma)$
of the local error growth being equal to $\gamma$
at time $t=M \Delta t$. 
The introduction of a p.d.f. $P_t(\gamma)$  is
possible under the ergodicity assumption
(for a review of the ergodic theory, see for example Halmos (1956)).
The
p.d.f. is then related to the existence of an attractor set,
described by an invariant measure, which can be operatively defined by
computing the average fraction of the time spent by the evolving
system
in any portion of the
attractor.
The quantity $\langle  R(t,0)^q \rangle$ can be expressed by the
integral:

\begin{equation}
\langle R(t,0)^q \rangle = \int P_t(\gamma) e^{\gamma q M \Delta t}
d\gamma.
\label{eq:rq}
\end{equation}

As already pointed out, only for $M\rightarrow \infty$ do we have 
$P_t(\gamma) \rightarrow \delta(\gamma-\lambda_1)$.
For finite but large $M$, the large deviations theory 
(for a general presentation see Varadhan (1984) and Ellis (1985); for 
a more physical treatment see Frisch (1995)) characterizes 
$P_t(\gamma)$ as: 

\begin{equation}
 P_t(\gamma)  \sim e^{-S(\gamma) M \Delta t}, 
 \label{eq:ans} 
\end{equation}
where $S(\gamma)$ is a concave function such that $S(\gamma) \geq 0$ 
and $S(\gamma)=0$ for $\gamma=\lambda_1$.
The large deviations theory holds, in its simplest form, for
independent random variables. Like the law of large numbers and the
central
limit theorem, the large deviations theory has extensions to random
variables
with correlations, when these correlations decrease sufficiently fast.
These are the conditions under which $P_t(\gamma)$ takes the form 
(\ref{eq:ans}).
In the language of large deviations theory, $S(\gamma)$ is called 
{\it Cramer function} or {\it Cramer entropy} (Mandelbrot 1991).
Different systems will have different p.d.f.'s (i.e., different
$S(\gamma)$). In the next part of this subsection we will show how the
explicit form of the $L(q)$ exponents can be obtained from the Cramer
function $S(\gamma)$ of the system. We will also show the general
relation between $L(q)$ and $\lambda_1$.

Inserting (\ref{eq:ans}) into (\ref{eq:rq}), 
by saddle-point integration one obtains:

\begin{equation}
\langle R (t,0)^q \rangle \sim \int  
e^{[\gamma q - S(\gamma)] M \Delta t} d\gamma \sim e^{ L(q)M \Delta
t},
\label{eq:rqbis}
\end{equation}
where 

\begin{equation}
L(q) = \sup_\gamma [q \gamma - S(\gamma)]
\label{eq:max}
\end{equation}
and $sup_{\gamma}[f(\gamma)]$ means the smallest upper bound
of the function $f(\gamma)$.

Since $S(\gamma)$ is a concave function, the  maximum
value of the convex function $q \gamma - S(\gamma)$ is unique and 
attained at the value $\gamma_q$ at which

\begin{equation}
\frac{d(q \gamma - S(\gamma))}{d\gamma}\left .
\right|_{\gamma=\gamma_q}=0
\qquad \mbox{or}\qquad 
\frac{dS(\gamma)}{d\gamma}\left . \right|_{\gamma=\gamma_q}= q.
\label{eq:dS}
\end{equation}
Thus, from (\ref{eq:max}) one obtains:

\begin{equation}
L(q)= q \gamma_q-S(\gamma_q)
\label{eq:Lq}
\end{equation}
and 

\begin{equation}
\frac{dL(q)}{dq} =\gamma_q+q \frac{d \gamma_q}{dq}- 
\frac{dS(\gamma_q)}{dq}=\gamma_q.
\label{eq:dL}
\end{equation}

Since $L(0)=0$, 
it  follows from (\ref{eq:Lq}) that $S(\gamma_0)=0$.
Thus, as an immediate consequence of the general properties of  
$S(\gamma)$ 
, one obtains  $\gamma_0=\lambda_1$.
From (\ref{eq:dL}) the following expression for the maximum Lyapunov
exponent
can be obtained: 

\begin{equation}
\lambda_1={{\frac{dL(q)}{dq}} \vert} _{q=0}.
\label{eq:dLbis}
\end{equation}

The quantities $\gamma_q$ are the characteristic exponents describing
the predictability fluctuations of the dynamical system: if $L(q)$ is
a linear function of $q$, they reduce to a single exponent, namely
$\lambda_1$ (see eq.~(\ref{eq:dL})). If $L(q)$ does
not follow the linear relationship $\lambda_1 q$, then the complete
set of
exponents $\gamma_q$ is needed to characterize predictability
fluctuations. 
In this case the system is {\em intermittent}.
The functional form of $L(q)$ depends  on the system, and
specific assumptions on the p.d.f. $P_t(\gamma)$ have to be made
in order to perform a quantitative analysis.

\subsubsection{The log-normal case}
There is no general theory about the 
shape of the $S(\gamma)$ function, as 
different systems may have different predictability fluctuations.
If correlations are
weak enough, central limit arguments can be applied and a Gaussian law
for $P_t(\gamma)$ may be assumed (Paladin and Vulpiani 1987).
Several numerical studies (see, for example, 
Benzi {\em et al.} 1985; Benzi and Carnevale 1989) 
show to what degree 
the log-normal hypothesis for the statistics of the error growth 
can reproduce the computed $L(q)$'s
for simple dynamical and atmospheric models. 
Let us recall here the main points which follow from this hypothesis,
assuming  a quadratic behavior for $S(\gamma)$:

\begin{equation}
S(\gamma)=(\gamma-\lambda_1)^2/2 \mu .
\label{eq:s}
\end{equation}

From eqs. (\ref{eq:ans}) and (\ref{eq:s}) it follows that the
normalized p.d.f. for $\gamma$ reads:

\begin{equation}
P_t(\gamma)=\frac{1}{\sqrt{2\pi\mu/t}}e^{-\frac{(\gamma-\lambda_1)^2}{
2\mu/t}}.
\label{eq:scemo}
\end{equation}

From the transformation rule
$P_t(R)=P_t(\gamma(R))\left|\frac{d\gamma}{dR}\right|$, 
the  log-normal distribution for the response function $R(t,0)$
follows as:

\begin{equation}
P_t(R) = \frac{1}{R \sqrt{2 \pi \mu t}}
 e^{-\frac{(\ln{R}-\lambda_1 t)^2}{2 \mu t}}.
\label{eq:p}
\end{equation}
The probability distribution is thus fully characterized by two 
parameters only:
\begin{equation}
\begin{array}{llll}
\lambda_1&=& \langle \ln{R(t,0)}  \rangle /t,\\
\mu&=&[\langle (\ln{R(t,0)})^2 \rangle - 
\langle \ln{R(t,0)} \rangle^2]/t, 
\end{array}
\label{mula}
\end{equation}
where $\lambda_1$ is the maximum Lyapunov exponent and 
$\mu $ is the second cumulant.

In the general case, the parameters $\lambda_1$ and $\mu$, as defined
in (\ref{mula}),
continue to be  important,
although they may not completely represent  the p.d.f.: 
they 
give respectively 
the mean value and the variance of the
$\gamma$-fluctuations.
For this reason, the $\mu$ parameter is usually referred to as 
the intermittency of the system.
The ratio $\mu/\lambda_1 = 1$ may be taken as 
 the borderline between weak and strong
intermittency. About the latter regime, an important point has to be
stressed.
Consider the most probable response function value $\tilde{R}$ 
(obtained by solving $\frac{dP_t(R)}{dR}=0$) and the 
mean value $\langle R \rangle$.
In the log-normal case their expressions read:

\begin{equation}
\tilde{R}= e^{\lambda_1 t (1-\mu/\lambda_1)}\qquad\mbox{and}\qquad
\langle R \rangle= 
e^{\lambda_1 t (1+\mu/(2\lambda_1) )}.
\label{eq:ud}
\end{equation}
 From eqs. (\ref{eq:ud}) it  follows that
a rough estimate of the response
that, for example, takes the most probable value $\tilde{R}$ as 
representative of the distribution, fully
breaks down for $\mu/\lambda_1 > 1$. This approximation
predicts a decreasing error for 
long times
 ($\lim_{t \rightarrow \infty}\tilde{R}=0$) 
instead of the chaotic error growth 
 characterized by a positive exponent 
($\lim_{t \rightarrow \infty} \langle R \rangle =\infty$). 

Using  $S(\gamma)$ given by (\ref{eq:s}) in the
expression (\ref{eq:dS}) we obtain:
\begin{equation}
\gamma_q=\lambda_1+\mu q
\label{eq:scemo2}
\end{equation}
which, substituted into (\ref{eq:Lq}), gives
the generalized Lyapunov exponents for the log-normal case:

\begin{equation}
L(q) = \lambda_1 q + \frac{1}{2} \mu q^2.
\label{eq:l_qua}
\end{equation}

As a reasonable definition of the characteristic predictability time
$T$
of the system,
one can  consider the following, 
which takes into account the effects of fluctuations:
\begin{equation}
T \sim \frac{1}{L(1)} \sim \frac{1}{\lambda_1+(1/2)\mu} .
\label{eq:predi}  
\end{equation}

It is reasonable to assume that, in the general case, $L(q)$
would be bounded between the linear shape (i.e., absence of
fluctuations) 
and the quadratic shape
(i.e., strong fluctuations).
In many cases, the log-normal shape is a good
approximation for small
deviations of $\gamma$ around its mean value, i.e., it reproduces
quite
well the smallest moments of the distribution.
On the other hand, for large $q$'s, the moments cannot be 
of the log-normal type (Paladin and Vulpiani 1987),
since
log-normal moments grow more than exponentially with
$q$. This implies that $\lim_{q \rightarrow \infty}\gamma_q$
is not finite.  
This limit is related to the fastest error growth $R^*$ in the system,
which must be finite for all physical systems.
The log-normal approximation fails to reproduce the tails of
the 
distribution $P_t(R)$, that is, the largest fluctuations.\\

\subsubsection{The log-Poisson case}
The physical constraint that the fastest error growth must be finite
forces one  to look for  parametrizations of $P_t(\gamma)$ other than
Gaussian.
For this purpose, let us introduce, for the first time to our
knowledge
in predictability theory, recent ideas developed in 
statistical treatment of fully developed turbulence (the basic
concepts
can be found in She and Waymire (1995)).
  
One way to consider alternative probability distributions is by
observing that 
the 
quantity: 

\begin{equation}
R(t^\prime,t)= \frac{|\delta{\bf x}(t^\prime)|}{|\delta{\bf x}(t)|}
\label{eq:Rtt}
\end{equation}
obeys the multiplicative rule:

\begin{equation}
R(t^\prime,t)= R(t^{\prime},t^{\prime\prime}) R(t^{\prime\prime},t),
\label{eq:Rmul}
\end{equation}
 for any $t^{\prime\prime}$.
Thus, one is led to consider all possible probability functions 
$P_t(\gamma)$ which
are left functionally invariant by the multiplicative transformation 
(\ref{eq:Rmul}).
We shall call  this class 
of probability functions covariant.\\
By making the (strong) assumption of weak correlation
between $R(t^{\prime},t^{\prime\prime})$ and $R(t^{\prime\prime},t)$,
 an important class of distributions turns out to be covariant:
this is the class of the {\it infinitively divisible distributions }
(IDD). The normal and the Poisson distributions are the most
popular examples of IDD (for a comprehensive list and demonstrations 
of the properties of IDD, see for example Doob (1990)). 

In the following we shall consider the Poisson distribution as 
the simplest example of a non-Gaussian IDD that gives a more suitable
description of strong fluctuations  than the Gaussian distribution.
As we shall see, it permits also to satisfy the physical 
constraint that the fastest error growth has to be finite.

Notice that this simple (and discrete) 
stochastic model is not intended to 
represent the p.d.f. of a generic (continuous) physical system. Our
aim 
here is twofold: on one hand, we show that
the Poisson distribution, in spite of its simplicity,  
shares some important properties with realistic
p.d.f.'s; on the other hand,  we show that statistical
quantities can be successfully fitted by using  the Poisson formulation.
In this sense, the Poisson assumption gives a simple
way  to characterize and
quantify some predictability properties.
 
Let us  rewrite the response function in terms of a Poisson random
variable $x$, factorizing its exponential time dependence:

\begin{equation}
R(t,0) \sim e^{a t} \beta^{x},
\label{eq:errorpoiss}
\end{equation}
where $\beta \leq 1$ and $x$ is a random variable 
that follows  the Poisson distribution:

\begin{equation}
P_t(x=n) = \frac{(b t)^n e^{-b t}}{n!}.
\label{eq:Ppoiss}
\end{equation}
Using eq.~(\ref{eq:Ppoiss}), the qth-order moment of the response
function
$R$ can be easily calculated:
\begin{equation}
\langle R^q\rangle = \sum_n\frac{(bt)^n e^{-bt}}{n!}e^{aqt}\beta^{qn}
= e^{(aq -b)t}\sum_n\frac{(bt\beta^q)^n}{n!}=e^{[aq -b(1-\beta^q)]t}
\label{pipponi}
\end{equation}
and the expression for the generalized Lyapunov exponents becomes:
\begin{equation}
L(q)=a q - b (1-\beta^q).
\label{eq:Lpoiss}
\end{equation}

From eqs.~(\ref{eq:dL}) and (\ref{eq:dLbis}) we also obtain
\begin{equation}
\gamma_q=a +b \beta^q \ln\beta \qquad \mbox{and}\qquad
\lambda_1=a+b\ln\beta.
\end{equation}
Notice that for large $q$'s the generalized Lyapunov exponents 
given by (\ref{eq:Lpoiss})
behave as $L(q) \sim aq-b$. The $a$ parameter is 
associated with the fastest error growth in the system, $R^*=e^{at}$,
which is finite.
Thus, the log-Poisson p.d.f. does not show the pathologies typical
of the log-normal distribution.

A comparison between  $R^*$ and   $ \langle R \rangle$ can give
a measure of intermittency, namely:
\begin{equation}
\frac{\ln R^*}{\ln \langle R \rangle} = \frac{a}{L(1)}\equiv
\tilde{a}.
\label{new_indic}
\end{equation}  

To summarize, 
the considerations outlined in this section
allow us to investigate predictability
properties at large $t$ (i.e., on the attractor set) by means of quantities 
such as $\lambda_1$, $\mu$ and $L(q)$.
In particular,
one can consider the ratio $\mu/\lambda_1$ (particularly important if
a log-normal hypothesis is made) or the ratio $\tilde{a}$ (in the case
of
log-Poisson statistics) as  measures of 
predictability fluctuations.

\subsection{Characterization of predictability fluctuations at short times}
\label{sec:short_t}
In this section we show, with the help of a simple example, how it is
possible to extract quantitative information about the predictability
fluctuations of a system in the super-exponential error growth regime.
Here and for the first time in a meteorological context,
we use
recent techniques employed in turbulence data analysis, which lead to
a generalization of the concepts previously reported  for the
asymptotic regime. Again, the simple Poisson assumption can be made
and
useful statistical quantities can be evaluated. 

We are here concerned with the response
function for a finite time interval. 
A statement about the choice  of initial conditions has thus to be
made.
In the following, we make the familiar
assumption of an homogeneous and isotropic probability distribution of
the disturbance vector in phase space, which was assumed originally by
Lorenz (1965).

\subsubsection{An example}
In order to discuss the predictability fluctuations
at short times, we will consider the  
Lorenz (1963) system  (hereafter L63).
For $r \sim 24.74$ (the critical value for the fixed points to be 
unstable) a linear behavior of $L(q)$ is found (see Benzi {\em et al.}
1985): 
the system does 
not show intermittency in the long times regime.
The same behavior does not hold for short times, which are  
relevant in meteorological applications.

This point can be detected in Fig.~2, where the second and sixth order
moments
of the response function $R$ are shown, as an example,
as a function of time for
$r=28$.
In both graphs, two temporal ranges are clearly detectable
(the noise in the signal is simply due to lack in statistics).
For $t \gtrsim  \tau = 1.25 $ there is evidence of an 
exponential error growth, whose exponent, in these log-linear plots,
is given by the slope of the straight line fitting the data:
the slopes are  $L(2)\sim 2\lambda_1\sim 1.6$ 
(see Fig.~2(a)) and $L(6)\sim 6\lambda_1=4.8$ (see Fig.~2(b)).
These values confirm the linear law $L(q)=q\lambda_1$ 
for the generalized Lyapunov exponents 
found by Benzi {\em et al.} (1985). 

At shorter times ($t\lesssim \tau $), 
the error grows more rapidly. Furthermore,
a zoom over $t\lesssim 0.2 $ reveals a non-exponential behavior of 
the response function.
This feature makes it difficult 
to extract informations from the data inside the range 
$0\leq t \lesssim \tau$.

Notice that the separation between the two temporal regimes takes
place at a time scale $\tau$ which can be identified to be $\tau \sim
1/\lambda_1$. This seems to be a  non-trivial result, even if more 
examples based on other models are necessary
to get the confidence that this property is  general and not merely 
a coincidence.

In order 
to determine the statistical properties of the error growth at small
$t$, we try to consider different representations of our data.
Let us consider, as an example, the log-log plot of
the sixth order moment
against the second order moment (Fig.~3). As one can see,
such representation reduces
the spreading of points with respect to Fig.~2, 
thus improving the quality of the linear fit.
Furthermore, 
a zoom corresponding to the temporal range $t\lesssim 0.2$ (inside box
in Fig.~3)
reveals the following key point:
a linear behavior now takes place from 
very early time and
a measure of intermittency now becomes possible.
The best fit for long times ($t 1/\lambda_1$) gives the ratio 
$L(6)/L(2)\sim 3$
(i.e., no intermittency is present in the long-time regime). 
For $t < 
1/\lambda_1$ this ratio gives the values $4.9$, indicating 
a nonlinear behavior of $L(q)$ and thus the intermittent nature of the
system for short times.

Fig.~3 represents one of the main results obtained in this paper. 

From this example we argue that it is possible to achieve a
quantitative 
measure of predictability fluctuations even at short time, 
when non-exponential growth rates are observed.

\subsubsection{The ESS analysis at short times}
The results of Fig.~3 suggest to try a generalization of the concepts 
reported in sec.~2.1. Accordingly, one can continue to assume
the covariance property (\ref{eq:Rmul}) to be 
valid even for a time-dependence 
of the response $R(t,0)$
different from the exponential one. 

Thus, the hypothesis of IDD's governing the error growth statistics
continue to hold in the transient regime.
As an example, let us consider again the Poisson case.
The temporal dependence $e^t$ may be replaced  by 
a more general function $g(t)$, such that $g(0)=1$ and
$\lim_{t\rightarrow \infty}
g(t)=e^t$. Relation (\ref{eq:errorpoiss}) becomes:

\begin{equation}
R(t,0) \sim g(t)^{a} \beta^{x},
\label{eq:errorpoiss_t}
\end{equation}
where $x$ is a random variable with a Poisson distribution:

\begin{equation}
P_t(x=n) = \frac{[b~ \ln g(t)]^n e^{-b~ \ln g(t)}}{n!}.
\label{eq:Ppoiss_t}
\end{equation}

From (\ref{eq:errorpoiss_t})
 and (\ref{eq:Ppoiss_t}), it follows that the statistical moments of
the response
function at short $t$ take the form:

\begin{equation}
\langle R(t,0)^q \rangle \sim e^{L(q)\ln g(t)},
\end{equation}
with $L(q)$ given by (\ref{eq:Lpoiss}). 

The results of Fig.~3 suggest that
one may extract information
about predictability fluctuations in the system by considering 
ratios like $\ln \langle R^q \rangle/\ln \langle R^p\rangle $, which
allow us
to eliminate the functional dependence on $g(t)$.
This is the basic idea of the {\it Extended Self Similarity} (ESS)
techniques recently employed in   
fully developed turbulence (Benzi {\em et al.} 1993;  
Benzi {\em et al.} 1995): 
they  concern the possibility 
of observing a  scaling behavior of 
turbulent velocity and energy dissipation
fields over  an
enlarged range of spatial scales and in systems and conditions where 
these scaling properties  are not evident with standard techniques.  

The above  simple considerations allow extending the notion of 
generalized Lyapunov exponents in a simple way. 
Accordingly, we can define the quantities:
\begin{equation}
\frac{\ln \langle R^q \rangle}{\ln \langle R \rangle} = 
\frac{L(q) \ln g(t)}{L(1) \ln g(t)}\equiv
\tilde{L}(q).
\end{equation}
The constraint $\tilde{L}(1)=1$ leaves only two free parameters,
$\tilde{a}=a/L(1)$ and $\tilde{b}=b/L(1)$.
As already noted (see Eq.~(\ref{new_indic})), the quantity $\tilde{a}$
has a straightforward meaning and can be considered as a useful
intermittency indicator. 

By measuring $\tilde{L}(q)$ exponents, it is thus possible to obtain
quantitative information about predictability properties even at short
times.

Let us conclude this section by
introducing another new indicator that we cinsider useful in studying

predictability fluctuations at short times.
We define the quantity
$\tilde{P}(t)$ as:

\begin{equation}
 \tilde{P}(t) = \int_{R\leq 1} P_t(R) dR,
\label{eq:Ptilde}
\end{equation}
where $P_t(R)$ is the p.d.f. of
the response function $R(t,0)$.

The quantity $\tilde{P}(t)$ is the probability of observing an error 
$\delta {\bf x}(t)$ smaller or equal to the initial error $\delta {\bf
x}(0)$.
By construction, we have $\tilde{P}(0)=1$, while in the limit of large
$t$,
$\tilde{P}(t) \rightarrow 0$.
If there are no fluctuations in the predictability exponents, one 
obtains $\tilde{P}(t)=0$ for every $t \neq 0$.
This is not the case for intermittent systems. \\
Let us compare
the cases of log-normal and log-Poisson distributions. 
For a log-normal distribution, $\tilde{P}(t)$ decreases monotonically 
toward zero. 
An easy way to see it is the following:
after substituting eq.(\ref{eq:p}) into eq.(\ref{eq:Ptilde}) and definining
the variable
\begin{equation}
 z = \frac{lnR-\lambda_1 t}{\sqrt{2\mu t}},
\label{eq:newvar}
\end{equation}
one obtains the following expression of eq.(\ref{eq:Ptilde}) for the log-normal
case:
\begin{equation}
 \tilde{P}(t) = \int_{-\infty}^{-\frac{\lambda_1}{\sqrt{2 \mu}}
\sqrt{t}} \,\,\, \frac{e^{-z^2}}{\sqrt \pi} \,\,\, dz.
\label{eq:Ptilde_norm}
\end{equation}
The dependence on $t$ is now only in the upper bound of the
integral: as time increases, the range of integration decreases, thus
making $\tilde{P}(t)$ a monotonically decreasing function of $t$.\\
On the other hand, the short-time behavior in the
log-Poisson case is completely different:
$\tilde{P}(t)$ grows as $(1-e^{b\ln g(t) })$, up to
a time  $\bar{t}$ such that $g(\bar{t})
=\beta^{-1/a}$. Then, for large $t$, it decays 
toward zero, following a typical step behavior induced by the
discrete
character of the Poisson distribution.

The above considerations
imply that in an intermittent 
system the probability of observing
an error smaller or equal to the initial error
can be finite.
Moreover, the behavior of $\tilde{P}(t)$ for small $t$ is
completely different in 
the two cases of log-Poisson and log-normal distributions.
 This effect is due to the 
 finite maximum growth rate of the 
 Poisson case: the finiteness of the parameter $a$ allows a growth of 
 $\tilde{P}(t)$ for $t \leq \bar{t}$.

 These considerations illustrate the
straightforward physical meaning
of the quantity $\tilde{P}(t)$
 in predictability studies concerning the short time range.
Moreover, a systematic, quantitative investigation 
may be performed at any time by evaluating the $\tilde{L}(q)$
exponents.

\section{Application to a low-order primitive equation
model}
\label{sec:Lorenz}

In order to apply the theory outlined in the previous section,
we have considered the low-order model for atmospheric flow
introduced by Lorenz (1980). 
This model represents a suitable truncation of  simplified 
primitive equations (hereafter PE) on $\beta$-channel.
As in Lorenz (1980), the model equations are:

\begin{eqnarray}
{a}_{i} \frac {d x_i}{d \tau}&=&
a_i b_i x_j x_k -c(a_i-a_k)x_j y_k +c(a_i-a_j) y_j x_k 
\label{pe1} \\
& &- 2 c^2 y_j y_k-\nu_0 {a_i}^2 x_i + a_i(y_i - z_i) \nonumber \\
{a}_{i} \frac {d y_i}{d \tau}&=&
-a_k b_k x_j y_k - a_j b_j y_j x_k + c(a_k -a_j) y_j y_k 
\label{pe2} \\
& &- a_i x_i - \nu_0 {a_i}^2 y_i \nonumber \\
\frac {d z_i}{d \tau}&=&
-b_k x_j(z_k-h_k)-b_j(z_j-h_j)x_k+c y_j(z_k-h_k)
\label{pe3} \\
& &-c(z_j-h_j)y_k +g_0 a_i x_i -
\kappa_0 a_i z_i + F_i \nonumber
\end{eqnarray}
where $x_i$, $y_i$ and $z_i$ are the coefficients of the velocity 
potential, streamfunction and height fields, respectively.
Here, $(i,j,k)$ stands for any permutation over $(1,2,3)$.
For our studies, we have  used physical parameter values which
correspond to large scale motion in
the mid-latitude atmosphere,
in agreement with previous 
works (Lorenz 1980; Gent and McWilliams 1982; Vautard and Legras 1986;
Lorenz 1986; Curry {\em et al.} 1995)

The evolution equations for an infinitesimal error have been obtained
by linearization of eqs.~(\ref{pe1}-\ref{pe3}).
For all simulations presented in this paper, the familiar assumption
of homogeneous and isotropic p.d.f. of the initial disturbance vector
in the phase space is made (Lorenz 1965).  The number of realizations
(i.e., different trajectories) considered is of the order of
$10^4$. For each realization of the ensemble, the initial error has
been randomly selected on the sphere centered at the initial point of
the trajectory.
The numerical integrations have been performed using a fourth-order
Runge-Kutta scheme.

In order to show the capability of the parameter $\mu/\lambda_1$
to capture the fluctuations in the predictability, let us consider the
three
different regimes of the PE model discussed by Krishnamurthy (1985),  
corresponding to three different values of the forcing parameter:
$F_1=0.1$,  $F_1=0.25$ and $F_1=0.3$. 
The  physical regimes associated with such values are the following
(for details, see Krishnamurthy (1985)):
for $F_1=0.1$ the attractor of the PE system is 
free from gravity waves; for $F_1=0.25$ the trajectories vary 
chaotically, with an almost periodic 
behavior interrupted irregularly by short periods characterized by
high frequency gravity waves;
finally, for $F_1=0.3$, gravity waves are present all the time.  

To evaluate the two parameters $\lambda_1$ and $\mu$,
defined in (\ref{mula}),  each realization (trajectory) 
has been made to last for a time $t$  of the order of 
a hundred days, large enough to apply 
all the considerations of sec.~\ref{sec:large_t}.

In Tab.~1, 
the values of the positive maximum  
Lyapunov exponent, the intermittency, the ratio $\mu/\lambda_1$,
and the predictability time $T$
for the PE system in the three 
aforementioned regimes are shown.
As already noted, the value of 
$\mu/\lambda_1=1$ could be considered as the 
borderline between weak and strong intermittency. 
We can see from Tab.~1
that three different values of $\mu/\lambda_1$ occur in
the three regimes.
The signature of the gravity wave activity, 
in the cases $F_1=0.25$ and $F_1=0.3$,
is emphasized by the enhancement of both $\lambda_1$ and $\mu$,
with respect to the case $F_1=0.1$. The predictability of the system
is thus reduced by  persistent gravity waves. It is also worth
noticing
that strong intermittency (i.e., $\mu/\lambda_1 >1 $)
is present only in the intermediate case $F_1=0.25$,
characterized by the appearance  of bursts of chaoticity
containing high frequency gravity waves. 

 For the regime $F_1=0.1$, we have also performed 
a systematic numerical 
investigation of the generalized Lyapunov exponents at long and short
times.
The logarithm of the sixth order moment
of the response function $R$ as a function of time is reported in
Fig.~4.
Two regions are clearly detectable.
For $t$ large enough ($t \gtrsim 1/\lambda_1 \sim 16 \,days$) the
points can be 
 fitted by a straight line, indicating an 
exponential error growth: the slope
of the fitting line  gives the sixth order generalized Lyapunov
exponent
$L(6)$ (the noise in the signal is simply due to lack in statistics).
At shorter times ($t \lesssim 1/\lambda_1 $), 
the error grows more rapidly; moreover,
a zoom over the first 5 days reveals a strongly nonlinear
behavior. As we shall see in the next section, this is mainly 
associated with the presence of transient gravity waves, whose
amplitude 
is significant, especially in very the first times of integration.
 
Notice that, as in the L63 model, the
characteristic time scale associated with a change in the error growth
regime is of the order of $1/\lambda_1$.

The use of ESS techniques allows us to
improve the quality of the fit in the long-time region
and to extract information
on generalized Lyapunov exponents even at small $t$. As an example,
 the sixth order moment
against the first order moment is  shown in Fig.~5 in a log-log plot.
We can see from 
the zoom over the 5 day temporal range 
that a linear behavior occurs from very early times.
We have then evaluated for 14 moments 
(q=0.2, 0.4, 0.6, 0.8, 1, 2,..., 10) 
the $\tilde{L}(q)$ exponents at short and long times,
by performing linear fits in the two temporal regions 
$0 < t \le 10 \,$ days  and $40 \le t \le 300 \,$ days, respectively.
The results are shown in Fig.~6 :
the points  
are well fitted by the log-Poisson formula, with parameters 
$\tilde{a}_{l}=1.2$, $\tilde{b}_{l}=0.3$ (for the long-time
$\tilde{L}(q)$
values)
and $\tilde{a}_{s}=4.5$, $\tilde{b}_{s}=4.3$ (for the short-time 
$\tilde{L}(q)$ values).
In particular, notice that  
$\tilde{a}_{s} > \tilde{a}_{l}$, as the comparison of the 
two slopes at high $q$'s shows.
This is a quantitative indication of the more intermittent 
nature of the system
for short times (see eq.~(\ref{new_indic})).

Focusing our attention 
on the short-time regime,
we have also evaluated the function
$\tilde{P}(t)$ defined in sec.~\ref{sec:short_t},
by counting the number of realizations 
(trajectories) having $R(t,0)<1$ at a 
certain time t:

\begin{equation}
\tilde{P}(t)=\frac{\it{N}_{R(t,0)<1}}{\it{N}_{tot}}.
\label{e7}
\end{equation}

Fig.~7 
shows the $\tilde{P}(t)$  
for the PE model.
%
Due to the appearance of 
transient gravity waves, 
a very strong loss of predictability occurs 
immediately after the beginning of the 
PE model integration:
$\tilde{P}(t)$ is close to $0$ and starts to grow 
until $\bar{t} \sim 5$ days, a behavior in qualitative agreement
with a  log-Poisson character of the  $R(t,0)$ p.d.f..
After about $10$ days, gravity waves become lost and only 
slow oscillations remain.

From the above results, it appears that the presence of gravity waves 
(on the attractor set as well as in the initial transient) has strong 
effects on  predictability.
In particular, the loss of predictability immediately after
the first times of the  model integration may be
relevant for practical purposes.
The final question we want to address is thus the following:
may the elimination of gravity waves 
produce  an enhanced short-time predictability?
We shall try to answer this question in the next section.

\section{ Predictability fluctuations and their relations with gravity
wave activity}
\label{sec:init}

Our first aim is to study how 
the predictability  fluctuations are 
influenced by the disappearance of gravity waves
in models not supporting such waves.

We shall focus our attention on Lorenz's (1980)  algorithm 
exploiting the complete separation between quasi-geostrophic and
gravity wave 
frequencies, and leading to the well-known superbalance equation,
defining
the {\it slow manifold} (Leith 1980; Lorenz 1980). 

Defining 
${\bf{X}}^G=(X^G_1,X^G_2,X^G_3,X^G_4,X^G_5,X^G_6)$, where 
$X^G_{i}=x_i$ and
$X^G_{i+3}=z_i-y_i$ for $i=1,2,3$, while  
${\bf{X}}^R=(y_1,y_2,y_3)$, 
Lorenz (1980) exploited the scale separation imposing
the condition $\frac{d^p{\bf{X}}^G_{p,k}}{dt^p}=0$
which can be solved iteratively 
for each $p$ by the Newton method (Lorenz 1980).

The low-order dynamical system obtained from the PE model by
calculating 
the ${\bf{X}}^G$ components  through Lorenz's algorithm
(with $p=1$) is called here
{\it SuperBalance Equation} (SBE) model. 

We apply the notions outlined 
in sec.~\ref{sec:large_t} 
and \ref{sec:short_t} to the PE and SBE models in order to quantify
the effects of 
gravity wave activity
on the predictability fluctuations.

The positive maximum  
Lyapunov exponent, the intermittency, the ratio $\mu/\lambda_1$ 
and predictability time $T$ (see eq.~(\ref{eq:predi}))
 are shown in Tab.~2 for 
the two models.
Actually, as we can see from this table, 
these asymptotic 
quantities do not permit us
to discriminate between the PE and SBE models. 
The very slight differences in these chaotic indicators confirm  
the capability of SBE to reproduce the asymptotic statistical
properties of the PE system. 
Let us then come back to the original question:
in what sense does elimination of transient gravity waves  make 
the SBE model less chaotic than the PE model?

An answer to this question can be given by exploiting
the short-time analysis presented in sec.~\ref{sec:short_t}.
In Fig.~8 
we report the same plot of Fig.~4, but for the SBE model.
As in the PE model, two regions are clearly detectable,
with the error growth for short times faster than that for long times.
However, the zoom over the first 5 days shows  an exponential
(that is linear in the graph) error growth since the very first time,
unlike the PE case.
This is due to  the projection onto the slow
manifold which causes the gravity waves to die out completely.

In order to improve the quality of the fit and  make
a comparison possible,
we perform for the moments of the error growth 
the same ESS analysis already presented
for the PE case. In particular, $\tilde{L}(q)$ exponents 
of the SBE model  in the long-time regime turn out to 
be indistinguishable from those of
the PE model  shown in Fig.~6 (white circles). This fact
confirms the success of the SBE model
in reproducing the asymptotic statistical properties of the 
Lorenz model.

Let us focus our attention on the short-time behavior.
The $\tilde{L}(q)$ exponents for the SBE model,
obtained from a linear fit at short time, are reported in Fig.~9.
In order to simplify the comparison, the corresponding curve for the
PE model  
(already reported in Fig.~6 (black circles))
is shown.
Several comments are worthwhile.
The log-Poisson fit for the SBE model gives 
$\tilde{a}_s=2.9$  and $\tilde{b}_s=2.4$: 
in particular, we stress 
that the slopes $\tilde{a}_s$ are  smaller than in the PE case.
Hence, disappearance of gravity waves in the SBE
model decreases temporal intermittency, damping predictability 
fluctuations in a sensible way.
 
The enhanced predictability during the first days 
becomes more evident  if one considers 
$\tilde{P}(t)$ behavior 
for the PE and SBE models, shown in 
Fig.~10. 
As we can see, the SBE curve strongly differs 
from that of the PE system during the  
first days.
Notice that the
$\tilde{P}(t)$ starts near $1$  for the SBE model. 
It then follows that 
the very strong loss of predictability (i.e., $\tilde{P}(t)$ close to
$0$)
immediately after 
the first times of 
the PE model integration 
is certainly due to
the transient gravity waves:   
for the SBE model, where gravity waves are not present, there is 
only a slight loss of predictability (i.e., $\tilde{P}(t)$ close to
$1$)
during the first few days.

After about $10$ days, gravity waves disappear in the PE system and 
the behavior of
$\tilde{P}(t)$ for the two models becomes quite similar. \\
From the behavior of both
 $\tilde{L}(q)$ and $\tilde{P}(t)$, 
we can conclude that the SBE model systematically 
produces at short times a more 
predictable flow with respect to the PE model.\\
To our knowledge, this is the first statistical assessment concerning
the effects introduced by 
initialization procedures
on the predictability properties of an atmospheric model.


\section{Conclusions}
\label{sec:conclu}

In the Introduction we have posed three different questions which we
believe 
are relevant for the predictability problem in atmospheric flows.

The answer to the first question we posed (how to describe
predictability fluctuations) can be given in terms of 
generalized Lyapunov exponents, $L(q)$'s, 
which represent the growth rate of the
moments of the response to an initial error.
The statistics of $R(t,0)$ are characterized by an infinite set of 
exponents  related to $\langle R(t,0)^q \rangle$. The non linearity 
of $L(q)$ {\it vs} $q$ reflects the intermittent
nature of the temporal  evolution of the error growth.
In this sense, this infinite set of exponents 
contains a level of  information
which is not present in  the maximum Lyapunov exponent $\lambda_1$, 
$\lambda_1$ being able to quantify only the average  predictability
properties of a system. The use of $L(q)$'s is not completely new. 
They have been
already introduced in the framework of atmospheric models by Benzi and
Carnevale (1989).
Regarding specific hypothesis on the  $R(t,0)$ p.d.f., they
considered the log-normal case. Here we have proposed the use of a
log-Poisson p.d.f. (recently introduced in the statistical theory 
of turbulence for the treatment of spatial intermittency, see for
example She and Waymire (1995)) to overcome some of the deficiencies of the
log-normal p.d.f.. 

We have also focused our attention on the  short-time predictability
properties which is a rather complex problem. 
The fact that the error growth need not be exponential for
short times makes the analysis  more difficult, introducing a complex
functional dependence of the response function 
on the time $t$, which can vary depending on the system.

Using recent  ideas  introduced and applied 
in the context of fully developed 
turbulence, namely the Extended Self Similarity techniques,
we have shown that the concepts of the 
 generalized Lyapunov exponents can be extended to 
the short-time regime,  introducing the 
$\tilde{L}(q)$ exponents. They are well-defined mathematical
quantities
that allow us to quantify the  intermittency in 
the system also at short times. 
This is one of  the major results achieved in this paper.

As a further measure of predictability fluctuations at short time, 
we have also considered the probability to have a decrease of the
initial 
error at time $t$ (what we called $\tilde{P}(t)$): this is a clear
indicator
concerning the possibility to have convergence of two trajectories 
at finite time,  due to large fluctuations around the average error
growth. 

We have applied these concepts to the case of a simple 
 atmospheric model (Lorenz, 1980), which 
 captures some features of synoptic scale motions at mid-latitudes.
We chose this model because its numerical study does not need the 
computational resources necessary in the case of
more complex and realistic models. Moreover, in spite  of its
simplicity, 
the model has a rich dynamics, characterized by appearance of the 
so-called gravity wave modes.  
This property has allowed us to answer the third question posed 
in the Introduction, that is, which is the connection, if any, between
short-time predictability fluctuations and initialization procedures.
To this end, we have considered the initialization procedure proposed
 by Lorenz (1980) leading to the SBE model.

Our results concerning the PE and SBE models can be summarized as
follows: 

\begin{itemize}

\item[-] {the generalized Lyapunov exponents  evaluated for the PE and
SBE models  
are well-reproduced by a statistical law 
of the log-Poisson type and differ both from  the non-intermittent
linear law
$L(q) \propto  q$ and the quadratic form implied by a log-normal
assumption;}

\item[-] {we have observed in the models two different regimes, 
associated with two temporal ranges:
in the short-time regime, the error grows more rapidly and
intermittently 
than in  the  asymptotic region. 
This means that standard (asymptotic) 
chaotic indicators as $\lambda_1$ and $\mu$ can be meaningless when 
considering short-time regimes;}

\item[-]{comparing  the asymptotic behaviors of the  models,
we conclude that the SBE procedure is successful in reproducing
the long-time statistics of the original PE model, since all the
asymptotic
chaotic 
indicators are practically indistinguishable in the two cases;}

\item[-]{elimination of gravity waves in the SBE model
strongly influences predictability
 for short times, giving a more predictable and less intermittent
flow.
The initialization procedure is then successful in decreasing 
the intermittency in the short
time region, which 
is strongly influenced by   transient  fast 
oscillations in the PE model;}

\item[-]{strongly non-exponential error growth was  observed
in the case of the model supporting gravity waves (PE)
after the first times of the
model integrations. This is the signature of
the  strong gravity wave activity.
Non-exponential behavior is strongly reduced 
in the SBE model, which does not support 
gravity waves. }

\end{itemize}
We would like to conclude with some remarks and suggestions for future
work.
It is necessary to better understand the physical meaning  both of
the transition
observed in the predictability properties 
(short-time  and long-time regimes) and 
the characteristic time at which this transition occurs.
This   characteristic time
seems to be strongly related to the maximum Lyapunov exponent. We 
 think that this feature (which we are currently investigating) 
is not peculia only to the models 
studied in this paper. 

Another point regards the study of the growth of non-infinitesimal
errors:
 indeed, in this paper we have considered the 
error growth in the so-called tangent space, that is, we have studied
 the behavior of an infinitesimal initial error, 
governed by the linearized version of the dynamical equations.
It would be interesting to perform a similar analysis for finite
initial errors,
in order to obtain more realistic information on the predictability 
when the error in the  initial conditions 
cannot be considered infinitesimal. 


Finally, we conclude with two remarks concerning the 
feasibility of our analysis  when more complex models involving
a large number of variables are considered. 
First, in our analysis we computed
the moments of the response function up to the tenth order just
to highlight the effect of predictability fluctuations. Actually,
these can
be captured also restricting the  analysis to lower order moments, 
thus  reducing the  related computational effort. 
Second, since the
predictability  is characterized by fluctuations,
the need for a good statistical ensemble of simulation runs is always
crucial.  This is independent  of the method used to perform the
statistical analysis and, in particular, is not peculiar to our
method.
The reason for this is that for systems characterized by  strong
intermittency, 
large fluctuations may occur with a non-vanishing probability
and their effects on the predictability have to be properly
accounted for in a statistical way. 
A large amount of statistics has thus to be gathered 
just in order to sample the tails of the p.d.f., where the rare events
are placed.  From this point of view, it is 
clear that over-simplified  atmospheric models suggest
a possible strategy to tackle the problem of
predictability fluctuations. 

\newpage
\section*{Acknowledgements}
We are grateful to L. Biferale, R. Festa, C.F. Ratto, O. Reale, 
M. Vergassola and A. Vulpiani for illuminating discussions, and Dr.
Claudio Paniconi for reviewing the text.
We thank the ``Meteo-Hydrological Center of Liguria Region''
where part of the numerical analysis was done. 
The author (2) acknowledge support from the Sardinian Regional
Authorities.
%
%
%

\newpage

\Large{{\bf References}}
\normalsize

\vskip 0.5cm


Benettin, G., L. Galgani, A. Giorgilli and J. M. Strelcyn,
		1980: Lyapunov characteristic exponents for smooth
dynamical 
systems; A method for computing all of them. Part I: Theory, and Part
II:
Numerical Application.  
		Meccanica {\bf15}, 9--20 and 21--30.\\

Benzi, R., G. Paladin, G. Parisi and A. Vulpiani, 1985:
Characterisation of intermittency in chaotic systems.
J. Phys. A, {\bf 18}, 2157--2165.   \\

Benzi, R., and G. Carnevale, 1989:
A possible Measure of Local Predictability.
J. Atmos. Sci., {\bf 46}, 3595--3598.	\\	

   Benzi, R., S. Ciliberto, R. Tripiccione, C. Baudet, 
		C. Massaioli, and S. Succi, 1993:
		Extended self--similarity in turbulent flows.
 		Phys. Rev. E, {\bf 48}, R29--R32.\\

 Benzi, R., S. Ciliberto, C. Baudet, and G. R. Chavarria, 1995:
              On the scaling of three--dimensionale homogeneous and
		isotropic turbulence.
 		Physica D, {\bf 80}, 385--398. \\



    Curry, J.H., S. E. Haupt, and  M. N. Limber, 1995:
Low--order models, initialization and slow manifold. 
                     Tellus, {\bf 47A}, 145--161.\\



 Doob, J.L., 1990: {\it Stochastic Processes.} 
		J. Wiley and Sons,  654 pp.\\

  Eckmann, J.P., and I. Procaccia, 1986:
		Fluctuations of dynamical scaling indices in nonlinear
		systems.
	       Phys. Rev. A, {\bf 34}, 659--661.\\

ECMWF, 1992: New developments in predictability. Workshop Proceedings,
13-15 November 1991.\\

 Ellis, R.S., 1985: 
		{\it Entropy, Large Deviations and Statistical
		Mechanics.} Springer, 364 pp.\\


Farrell, B. 1985: Transient growth of damped baroclinic
waves. J. Atmos. Sci., {\bf42},2718-2727.\\



 Frisch, U., 1995: {\it Turbulence}, Cambridge Univ. Press, 296 pp.\\

  Fujisaka, H., 1983: Statistical dynamics generated
		 by fluctuations of local Lyapunov exponents.
		Progr. Theor. Phys., {\bf 70}, 1264--1275.\\	

   Gent, P.R., and  J. C. McWilliams, 1982: Intermediate Model 
	Solutions to the Lorenz Equations: 
	Strange Attractors and Other Phenomena.
	J. Atmos. Sci., {\bf 39}, 3--13.\\

  Halmos, P.R., 1956:  {\it Lectures on Ergodic Theory}, 
                   Chelsea, 302 pp.\\


  Krishnamurthy, V., 1985: 
The slow--manifold and the persisting gravity waves.
PHD Thesis, MIT, 146 pp.\\

Lacarra, J. and O. Talagrand, 1988: Short range evolution of small
perturbations in a barotropic model. Tellus, {\bf40}A, 81-95.\\


 Leith, C.E., 1980: Nonlinear Normal Mode Initialization 
and Quasi--Geostrophic Theory. 
J. Atmos. Sci., {\bf 37}, 958--968.\\


Lorenz, E.N., 1963: Deterministic non--periodic flow.
J. Atmos. Sci., {\bf 20}, 130--141.\\

Lorenz, E.N., 1965: A study of the predictability of a 28-variable
atmospheric model. Tellus {\bf17}, 321-333.\\

 Lorenz, E.N., 1980: Attractor sets and Quasi--Geostrophic 
Equilibrium. 
J. Atmos. Sci., {\bf 37}, 1685--1699.\\

   Lorenz, E.N., 1986: On the existence of a slow manifold.
                     J. Atmos. Sci., {\bf 43}, 1547--1557.\\


Mandelbrot, B., 1991: Random multifractals: negative dimensions and
the resulting limitations of the thermodynamic formalism.
Proc. R. Soc. Lond., A {\bf 434}, 79--88.\\


 Molteni, F., R. Buizza, T.N. Palmer, and T. Petroliagis,
1996: The ECMWF Ensemble Prediction System: Methodology and
validation. 
Q.J.R. Meteor. Soc., {\bf 122}, 73--119.\\

Mukougawa, H., M. Kimoto, and S. Yoden, 1991:
A relationship between local error growth and quasi--geostrophic
states: 
case study in the Lorenz system.
J. Atmos. Sci., {\bf 48}, 1231--1237.\\

Mureau, R., F. Molteni, and T.N. Palmer, 1993:
Ensemble prediction using dynamically conditioned perturbations.
Q.J.R. Meteor. Soc., {\bf 119}, 299--323.\\

Nicolis, C., 1992: Probabilistic aspects of error growth in
atmospheric dynamics. Quart. J. Roy. Meteor. Soc., {\bf 118},
553-568.\\

 Oseledec, V.I., 1968: A multiplicative ergodic theorem. 
		Lyapunov characteristic numbers for dynamical systems. 
		Trans. Moscow Math. Soc., {\bf 19}, 197--231.\\

  Paladin, G., L. Peliti and A. Vulpiani, 1986:
     	      Intermittency as multifractality in history space. 
	      J. Phys. A, {\bf 19}, L991--L996.	\\	

 Paladin, G., and A. Vulpiani, 1987: Anomalous scaling laws in 
multifractal object.
Phys. Rep., {\bf 156}, 147--225.\\

 Palmer, T.N., R. Mureau and F. Molteni, 1990: The Montecarlo
 forecast. Weather, {\bf 45}, 198-207.\\

 She, Z.S., and E. C. Waymire, 1995:
		Quantized energy cascade and log--Poisson statistics 
		in fully developed turbulence.
		Phys. Rev. Lett., {\bf 74},
	       262--265.\\




Trevisan A., 1993: Impact of transient error growth on global average
predictability measures. J. Atmos. Sci., {\bf 50}, 1016-1028. \\

Trevisan A., and R. Legnani, 1995: Transient error growth and local
predictability: a study in the Lorenz system. Tellus, {\bf 47}A,
103-117.\\


 Varadhan, S.R.S., 1984: {\it An Introduction to the Theory 
               of Large Deviations.}
              Springer--Verlag, 196 pp.\\

 Vautard, R., and  B. Legras, 1986: 
Invariant Manifold, Quasi--Geostrophy and Initialization.  
              J. Atmos. Sci., {\bf 43}, 565--584.\\


\newpage

\begin{table}[p]
\begin{center}
\begin{tabular}{|c|c|c|c|c|}
\hline
$F_1$ & $\lambda_1 (days^{-1})$ & $\mu (days^{-1})$ & $\mu/\lambda_1$ 
& Predictability $(days)$ \\
\hline
$0.1$  & $6.3\times 10^{-2}$  & $2.5 \times 10^{-2}$ & $0.4$ & $13.2$\\
$0.25$ & $1.7 \times 10^{-1}$  &$2.8\times 10^{-1}$ & $1.6$ & $3.2$ \\
$0.3$  & $4.2 \times 10^{-1}$  &$3.1\times 10^{-1}$ & $0.7$ & $1.7$\\
\hline
\end{tabular}
\end{center}
\caption{Maximum Lyapunov exponent $\lambda_1$, intermittency
$\mu$, 
their ratio $\mu/\lambda_1$
and predictability for the PE model for different values of the 
forcing parameter.}
\end{table}

\begin{table}[p]
\begin{center}
\begin{tabular}{|c|c|c|c|c|}
\hline
Models & $\lambda_1 (days^{-1})$ & $\mu (days^{-1})$ & $\mu/\lambda_1$
 & Predictability $(days)$ \\
\hline
PE   & $6.3 \times 10^{-2}$  & $2.5 \times 10^{-2}$ & $0.4$ & 13.2\\
SBE  & $6.5 \times 10^{-2}$ & $2.0 \times  10^{-2}$ & 0.3 & 13.3 \\
\hline
\end{tabular}
\end{center}
\caption{Maximum Lyapunov exponent $\lambda_1$, intermittency
$\mu$, 
their ratio $\mu/\lambda_1$
and predictability for the PE and SBE  models at $F_1=0.1$.}
\end{table}

\newpage
\mbox{}

\begin{figure}
\epsfclipon
\epsfysize=8 cm
\epsfxsize=8 cm
\centerline{\epsffile{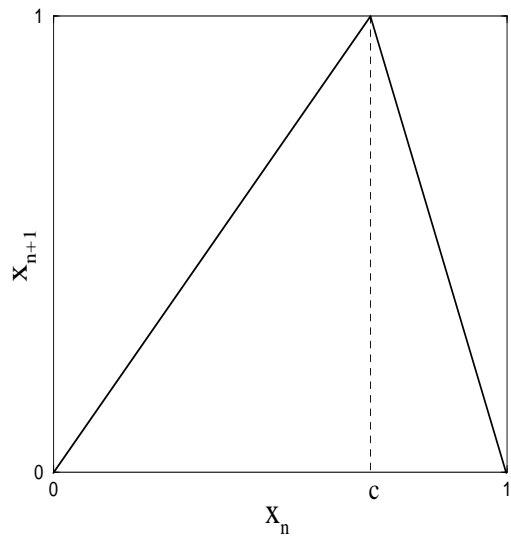}}
\caption{The one-dimensional ``tent map''.}
\end{figure}

\mbox{}
\newpage

\begin{figure}
\epsfclipon
\epsfysize=10 cm
\epsfxsize=10 cm
\centerline{\epsffile{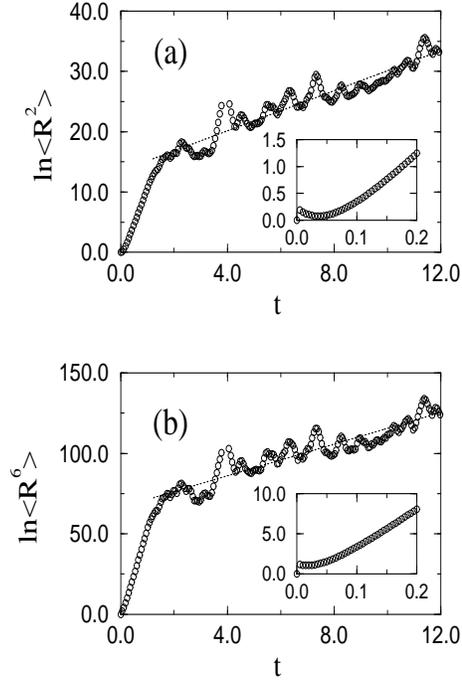}}
\caption{The logarithms of the second (a) and sixth (b) 
order  moments
of the response function $R(t,0)$ as  functions of time $t$ for the
L63 model.
Dotted lines result from linear fits of the data for $t >
1/\lambda_1$: 
their slopes are $\sim 2 \lambda_1$ and $\sim 6 \lambda_1$,
respectively.
Inside boxes are  zooms over the very short-time behaviors.}
\end{figure}

\mbox{}
\newpage

\begin{figure}
\epsfclipon
\epsfysize=8 cm
\epsfxsize=10 cm
\centerline{\epsffile{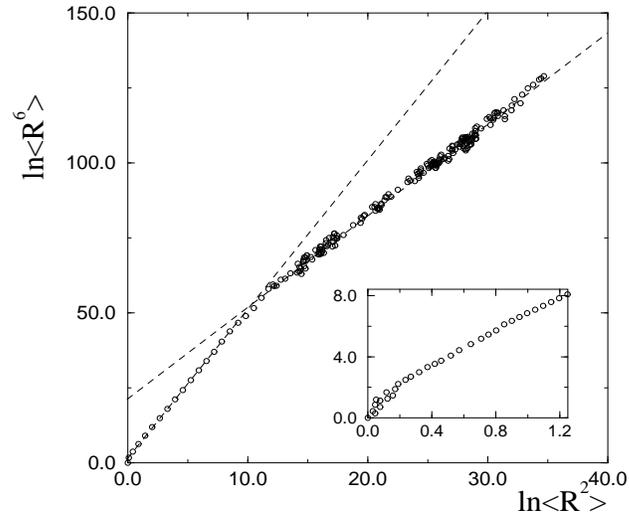}}
\caption{Log-log plot of the sixth order  moment 
against the second order  moment of the response function $R(t,0)$
for the L63 model.
Dashed lines result from linear fits of the data in the short and long
time regimes.
Inside box is a zoom over the very short-time behavior.}
\end{figure}

\mbox{}
\newpage

\begin{figure}
\epsfclipon
\epsfysize=8 cm
\epsfxsize=10 cm
\centerline{\epsffile{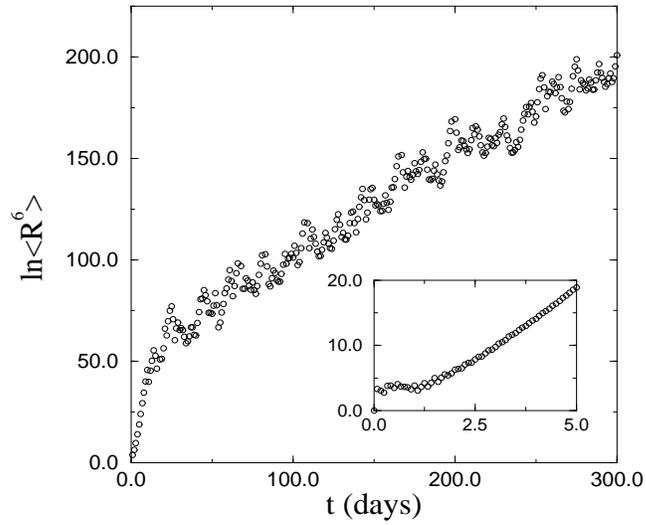}}
\caption{The logarithm of the sixth order  moment
of the response function $R(t,0)$ as a function of time $t$ for the PE
model.
Inside box is a zoom on the first 5 days behavior.}
\end{figure}

\mbox{}
\newpage

\begin{figure}
\epsfclipon
\epsfysize=8 cm
\epsfxsize=10 cm
\centerline{\epsffile{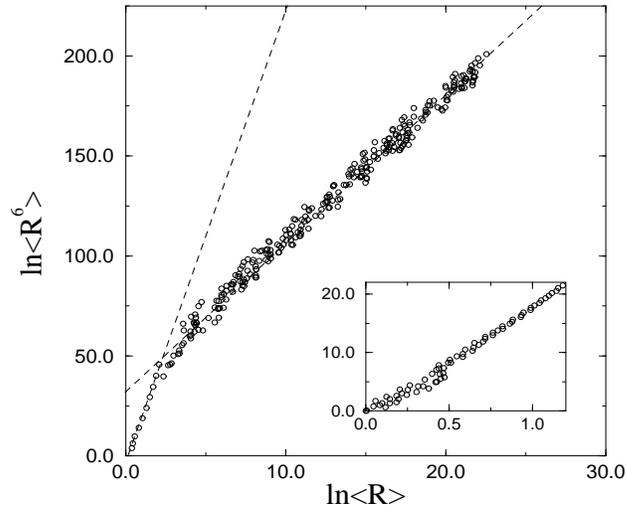}}
\caption{Log-log plot of the sixth order  moment 
against the first order  moment of the response function $R(t,0)$
for the PE model.
Dashed lines result from linear fits of the data in the short and long
time regimes.
Inside box is a zoom on the first 5 days behavior.}
\end{figure}

\mbox{}
\newpage

\begin{figure}
\epsfclipon
\epsfysize=8 cm
\epsfxsize=10 cm
\centerline{\epsffile{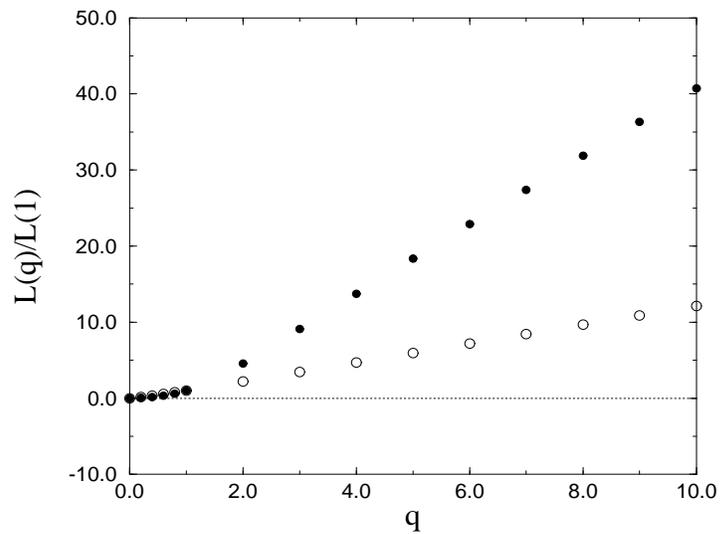}}
\caption{The $\tilde{L}(q)$ exponents as a function of $q$ 
for the PE model: black circles are the values for $0 < t \le 10 \,$
days,
while white circles denote the long--time values ($40 \le t \le 300 \,
days $).}
\end{figure}

\mbox{}
\newpage

\begin{figure}
\epsfclipon
\epsfysize=8 cm
\epsfxsize=10 cm
\centerline{\epsffile{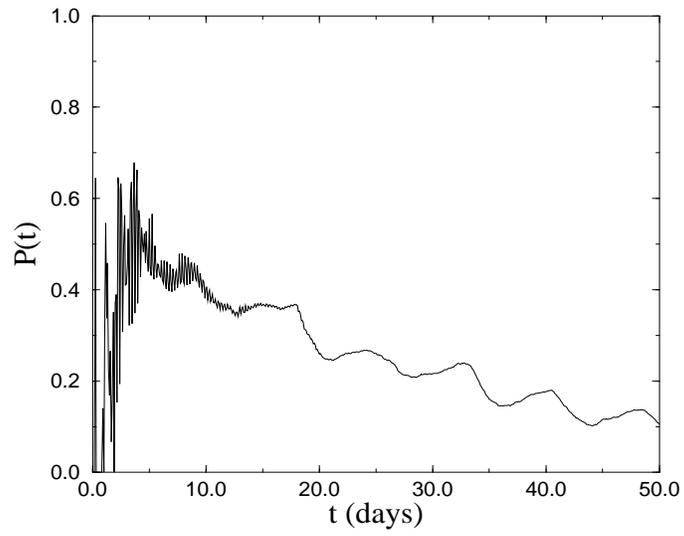}}
\caption{Time behavior of $\tilde{P}(t)$ for the PE model.}
\end{figure}

\mbox{}
\newpage

\begin{figure}
\epsfclipon
\epsfysize=8 cm
\epsfxsize=10 cm
\centerline{\epsffile{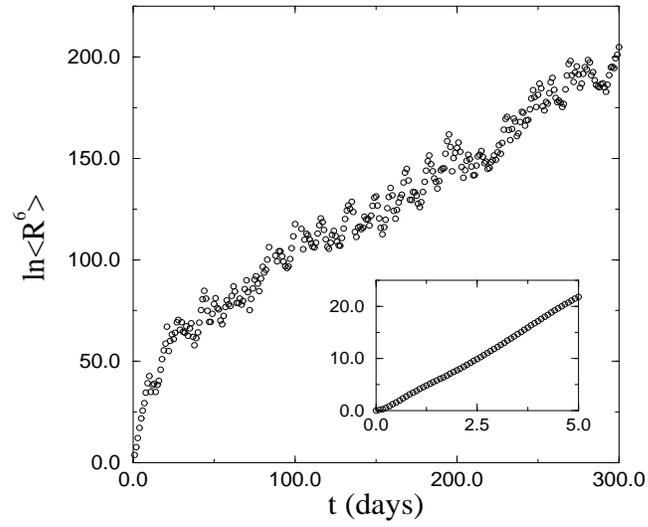}}
\caption{The same as in Fig.~4, but for the SBE model.}
\end{figure}

\mbox{}
\newpage

\begin{figure}
\epsfclipon
\epsfysize=8 cm
\epsfxsize=10 cm
\centerline{\epsffile{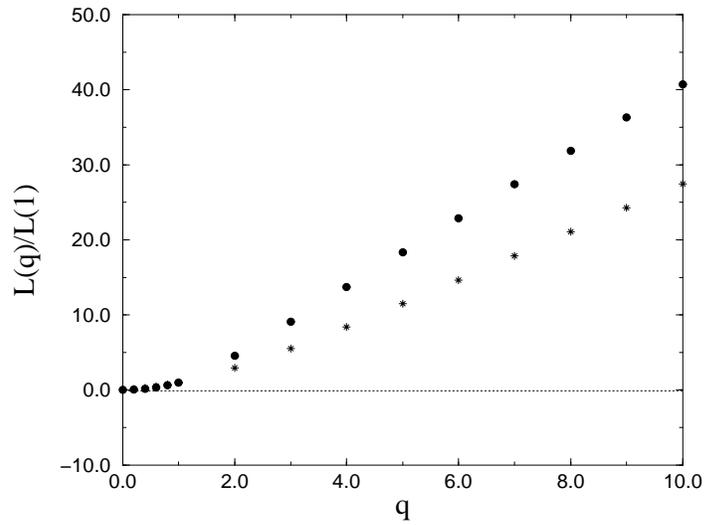}}
\caption{The $\tilde{L}(q)$ exponents as a function of $q$ 
for the PE and SBE models in the short--time regime ($0 < t \le 10\; 
days$):
circles and stars are the 
PE and SBE values, respectively.}
\end{figure}

\mbox{}
\newpage

\begin{figure}
\epsfclipon
\epsfysize=8 cm
\epsfxsize=10 cm
\centerline{\epsffile{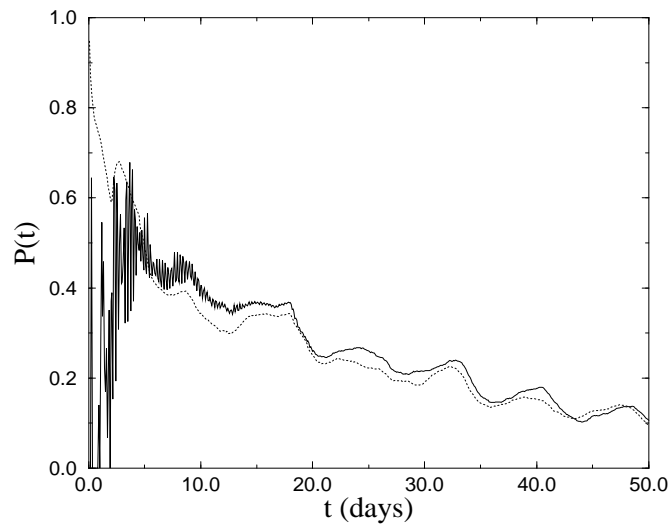}}
\caption{Time behavior of $\tilde{P}(t)$ for PE (thin curve)
and SBE (dotted curve) models.}
\end{figure}

%
\end{document}